\documentclass[runningheads]{llncs}
\usepackage{graphicx}
\begin{document}
\title{Job Offers Classifier using Neural Networks and Oversampling Methods}
%
%
\author{Germán Ortiz\inst{1}\orcidID{0000-0003-0655-5667} \and
Gemma Bel Enguix\inst{2}\orcidID{0000-0002-1411-5736} \and
Helena Gómez-Adorno\inst{3}\orcidID{0000-0002-6966-9912}\and
Iqra Ameer\inst{4}\orcidID{0000-0002-1134-9713}\and
Grigori Sidorov\inst{4}\orcidID{0000-0003-3901-3522}
}
\authorrunning{G. Ortiz et al.}
%

\institute{Posgrado en Ciencia e Ingeniería de la Computación\\ Universidad Nacional Autónoma de México, Mexico City, Mexico\\
\email{jortizb@iingen.unam.mx}\and
Instituto de Ingeniería, Universidad Nacional Autónoma de México, Mexico City, Mexico\\
\email{gbele@iingen.unam.mx} \and
Instituto de Investigaciones en Matemáticas Aplicadas y en Sistemas, Universidad Nacional Autónoma de México, Mexico City, Mexico\\
\email{helena.gomez@iimas.unam.mx} \and
Instituto Politécnico Nacional, Centro de Investigación en Computación,
Mexico City, Mexico\\
\email{iqra@nlp.cic.ipn.mx\\ sidorov@cic.ipn.mx}
}

%
\maketitle              
\begin{abstract}
Both policy and research benefit from a better understanding of individuals' jobs. However, as large-scale administrative records are increasingly employed to represent labor market activity, new automatic methods to classify jobs will become necessary. We developed an automatic job offers classifier using a dataset collected from the largest job bank of Mexico known as \textit {Bumeran}\footnote{\url{https://www.bumeran.com.mx/} Last visited: 19-01-2022.}. We applied machine learning algorithms such as Support Vector Machines, Naive-Bayes, Logistic Regression, Random Forest, and deep learning Long-Short Term Memory (LSTM). Using these algorithms, we trained multi-class models to classify job offers in one of the 23 classes (not uniformly distributed): Sales, Administration, Call Center, Technology, Trades, Human Resources, Logistics, Marketing, Health, Gastronomy, Financing, Secretary, Production, Engineering, Education, Design, Legal, Construction, Insurance, Communication, Management, Foreign Trade, and Mining. We used the SMOTE, Geometric-SMOTE, and ADASYN synthetic oversampling algorithms to handle imbalanced classes. The proposed convolutional neural network architecture achieved the best results when applied the Geometric-SMOTE algorithm.

\end{abstract}
\section{Introduction}
Mexico City has around 9 million inhabitants and is the state with the highest population density in the entire country. During the last quarter of 2018, there were 4.5 million economically active inhabitants, of which 230 thousand were unemployed \footnote{INEGI. Encuesta Nacional de Ocupación y Empleo (ENOE) \url{https://www.inegi.org.mx/programas/enoe/15ymas/default.html\#Tabulados} Last visited: 19-02-2022.}.


One of the biggest problems between supply and demand in the labor market is the asymmetry of information, which decreases the efficiency in the search for job opportunities~\cite{hart1983optimal}.

In this context, the Secretary of Labor and Employment Promotion (STyFE, \textit{Secretaria de Trabajo y Fomento al Empleo}), developed the project Diagnosis of Skills in Demand  (DiCoDE, \textit{Diagnóstico de Competencias Demandadas}), a system that monitors the job offers displayed on job portals in Mexico City to perform analysis on them.
Our work focuses on improving the functioning of the described system through the automatic classification of published offers in order to identify job profiles, skills and competencies demanded by employers and thus reduced the asymmetry mentioned above.

In this article, we develop two proposals for automatic classifiers of job offers for DiCoDE using two neural network architectures, a recurrent neural network and a convolutional neural network, in which we face the problem of unbalanced data set. A dataset is unbalanced if the number of instances of one or more classes is significantly higher or lower relative to the other~\cite{more2016survey} classes. The problem of class imbalance has gained more interest in recent years~\cite{imbalancedClassification} due to the growth of the field of machine learning applications, such as sentiment analysis \cite{9682721}, disease detection, and fraudulent phone call detection, among others. Since these data sets represent real-world problems, it is impossible to have the same number of examples of each of the classes involved. One of the ways to solve this problem is oversampling, which consists of increasing the number of examples of the minority classes, either by replicating some instances of it or by generating new examples using some criterion, such as the k- nearest neighbors algorithm \cite{imbalancedTSProblem}.

In this paper, we present a series of experiments using various oversampling techniques to improve the performance of classifiers using \cite{evaluationMetrics} Accuracy, Recall, Precision, and F1 as evaluation metrics.
To balance the class distribution of our dataset, we use the SMOTE~\cite{Chawla02smote:synthetic} technique, the Geometric-SMOTE~\cite{geometricSMOTE} variant, and the ADASYN~\cite{adasyn} algorithm.

The content of the article is organized as follows; in section \ref{sec:trabajoRelacionado} we describe related work to the oversampling methods used in our classifier proposals, as well as the use of neural networks in multiclass classification tasks. In section \ref{sec:Dataset}, we detail the characteristics of the corpus used for training and testing the neural networks. Subsequently, we address the methodology of the experiments carried out, in which the operation of the SMOTE, Geometric-SMOTE, and ADASYN algorithms is explained. In addition, we describe the pre-processing carried out on the text, the proposed neural network architectures, the training parameters used, and additional experiments using traditional machine learning algorithms to compare their performance with our proposals. In section \ref{sec:Resultados}, we show the results obtained and their comparison; Finally, in section \ref{sec:conclusiones}, we offer the conclusions of the work carried out and the direction of future work.


\section{Related Work}\label{sec:trabajoRelacionado}

In recent years, the problem of dealing with unbalanced data sets has gained more attention from the scientific community.

Regarding the generation of new examples of minority classes, \textit{Mohasseb et al.,} \cite{questionClasification}, propose the hierarchical-SMOTE algorithm to deal with imbalanced data in the Questions Classification task. This algorithm creates a grammar pattern to analyze each class in the dataset, then SMOTE algorithm is applied over all minority classes. The authors use a dataset that contains 1160 questions distributed in 6 classes and use a Naive Bayes classifier. The evaluation metrics they use are Precision, Recall, and F1, whose best results are 0.851, 0.865, and 0.847, respectively.

On the other hand, \textit{Douzas et al.,} \cite{landCoverClassification} face the unbalanced learning problem in the land cover classification task, in which a data set composed of 1694 examples is distributed in 8 classes. The class with the largest number of examples has 761 instances, while the category with the fewest number of elements has only 4 elements. In a matter of experiments, the evaluation metrics Accuracy and F1 and geometric mean are used. The algorithms used as classifiers are Logistic Regression, K-nearest neighbors, decision trees, Gradient Boosting and Random Forest. The Geometric-SMOTE algorithm achieves the best results in terms of F1 and geometric mean in the 4 algorithms, with Random Forest being the highest values with 0.341 and 0.572, respectively.

In the field of the telecommunications industry, \textit{Aditsania et al.,}~\cite{churnPrediction} deal with the problem of customer loss by developing a classifier that allows determining whether a customer is likely to stop using a service or not. To carry out this task, a data set belonging to an Indonesian telecommunications company was used, made up of 200,387 examples with 55 characteristics each, in which only 4\% correspond to examples of clients labeled as prone to stop using the service. Counteracting the imbalance of the data set, the ADASYN algorithm was used. A multilayer neural network trained using the backpropagation algorithm was used as a classifier. The results of the experiments show an interesting behavior since the highest Accuracy value is reached when the oversampling algorithm is not used. However, for this case, the F1 value is equal to zero, which implies that the system can only predict the majority class. With the implementation of the ADASYN algorithm, the values of Accuracy and Value-F are 0.93 and 0.46, respectively.

In the field of neural networks, \textit{Nowak et al.,} \cite{lstm} present a short text classifier using recurrent neural networks, specifically of the Long Short Term Memory (LSTM) type. 3 data sets were used to carry out their experiments, the first of them on messages classified as spam and non-spam with a distribution of 13.4\% and 86.6\%, respectively. The second one is about advertising, made up of 53.3\% of examples considered accepted and 46.7\% considered messages rejected. The last data set used consisted of book reviews classified as negative, positive, and neutral. As part of the pre-processing of the text, the text was converted to lowercase, punctuation marks and special characters were removed, numbers were removed, and a dictionary of unique words was created for each corpus. The evaluation was made based on the Accuracy obtaining values of 99.798\%, 94.497\%, 84.415\%, respectively.

In the same field, \textit{Lai et al.}~\cite{recurrentConv} present a classifier model based on a recurrent and convolutional neural network architecture to capture contextual information during the learning of word representations, as well as identify which words have a key role in the texts. To carry out their experiments, 4 data sets were used, the first of them: \textit{20newsgroup}, of which only 4 classes of 20 available are considered; the second: \textit{fudan}, composed of texts in Chinese distributed in 20 classes; third: \textit{ACL Anthology network}, made up of scientific papers divided into 5 classes and finally, the fourth data set: \textit{Stanford Sentiment Treebank}, made up of movie reviews in 5 classes. For performance evaluation, the first data set is evaluated using F1; the remaining sets use Accuracy as the evaluation metric. \textit{word embeddings} were used as word representations by the Skip-gram algorithm using Wikipedia for both Chinese and English. Additionally, comparisons were made using traditional machine learning methods and various neural network architectures. The model performs best on 3 of the 4 data sets, with an F1 of 96.49 for the \textit{20newsgroup} set, and Accuracy of 95.20 and 49.19 for the \textit{fudan} and \textit{ACL sets. Anthology network}, respectively.

\section{Dataset}\label{sec:Dataset}
Our dataset consists of 979,956 job offers written in Spanish, each of them belonging to one of the following 23 categories: Sales, Administration, Call Center, Technology, Trades, Human Resources, Logistics, Marketing, Health, Gastronomy, Financing, Secretary, Production, Engineering, Education, Design, Legal, Construction, Insurance, Communication, Management, Foreign Trade, and Mining. The job offers, as well as the categories used, were collected from the website Bumeran. Using a web scraping program \cite{webScraping} written in the R programming language. After performing the preprocessing described in \ref{subsec:Preproc} the distribution of the dataset is shown in Table \ref{tab:estadisticasDataset}.

\begin{table}[ht!]
    \caption{Dataset statistics after eliminating equal job offers}
    \centering
    \begin{tabular}{|c|c|c|}
        \hline
       Category & Instances & Percentage \\
       \hline
       Sales & 13,002 & 22.59\%\\
        Administration & 8,730 & 15.16\%\\
        Call center & 8,453 & 14.68\%\\
        Technology & 5,559 & 9.65\% \\
        Trades & 3,973 & 6.90\% \\
        Human Resources & 2,359 & 4.10\% \\
        Logistics & 2,206 & 3.83\% \\
        Marketing & 1,663 & 2.89\% \\
        Health & 1,610 & 2.80\% \\
        Gastronomy & 1,343 & 2.33\% \\
        Financing & 1,267 & 2.20\% \\
        Secretary & 1,236 & 2.15\% \\
        Production & 1,129 & 1.96\% \\
        Engineering & 881 & 1.53\% \\
        Education & 702 & 1.22\% \\
        Design & 661 & 1.15\% \\
        Legal & 645 & 1.12\% \\
        Construction & 622 & 1.08\% \\
        Insurance & 573 & 0.99\% \\
        Communication & 417 & 0.72\% \\
        Management & 272 & 0.47\% \\
        Foreign Trade & 228 & 0.40\% \\
        Mining & 41 & 0.07\% \\
       \hline
       Total & 57,572 &\\   
       \hline
    \end{tabular}
    \label{tab:estadisticasDataset}
\end{table}

\section{Methodology}\label{sec:Metodologia}

This section briefly describes the oversampling algorithms used in this work. We explain the pre-processing performed on the text, the neural networks architectures used to perform classification as well as the parameters used to carry out the experiments.

\subsection{Pre-processing}\label{subsec:Preproc}
Text Pre-processing was the following: first, the text was converted to lower-case, then punctuation marks were eliminated as well as special characters such as tab character and new-line symbol. Finally, we delete Spanish stopwords using the NLTK library \cite{nltk,ameer2019cic}.

After performing text pre-processing, we searched to eliminate repeated job offers to avoid overfitting our neural network architecture. Table \ref{tab:estadisticasDataset} shows the distribution of samples after the removal of equal job offers.

\subsection{Oversampling Algorithms}\label{subsec:Oversampling}
Unlike random oversampling, in which examples belonging to the minority class are randomly selected to be replicated and added to the training dataset, in our work, we use synthetic oversampling algorithms, where new examples are generated based on a series of criteria and using the \textit{K} nearest neighbors algorithm. We use the following algorithms:

\begin{itemize}

\item SMOTE: The Synthetic Minority Over-sampling Technique (SMOTE) algorithm generates new instances of the minority class by selecting instances of it, then find \textit{K} nearest neighbors to that instance and synthetically generates new instances. These generated instances are located along the line segment between the real selected instance, and its \textit{K} nearest neighbor \cite{Chawla02smote:synthetic}.

\item Geometric-SMOTE: The geometric-SMOTE algorithm generates new instances of the minority class based on its \textit{K} nearest neighbors the same way as SMOTE algorithm, but instead of using a line segment,  it uses a geometric region, which is usually a hyper-spheroid \cite {geometricSMOTE}.

\item ADASYN: The key idea behind this algorithm is to use a density distribution as a criterion to automatically decide the number of synthetic samples to generate for each minority sample, adaptively changing the weights of different minority samples to compensate for skewed distributions \cite{learning_imbalanced_data}.

\end{itemize}

We use the Python implementation of these algorithms available in the imbalanced-learn library \cite{DBLP:journals/corr/LemaitreNA16}.

\subsection{Classification}\label{subsec:Clasificador}
\begin{figure}[ht!]
    \centering
    \includegraphics[width=4.0cm]{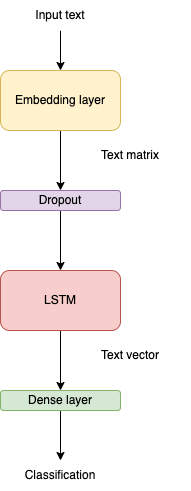}
    \caption{Block diagram for the proposed recurrent neural network architecture (LSTM)}
    \label{fig:arquitecturaClasificador}
\end{figure}

We built two different neural network architectures; the first architecture has an embedding layer to get a matrix representation of each text. We initialize the weights in this layer using the word2vec algorithm trained on the Spanish Billion Word Corpus \cite{cardellinoSBWCE}. This algorithm gets word vector representations on a continuous vector space whose objective is to preserve the semantic and syntactic similarity between the words \cite{ghannay-etal-2016-word}. After the embedding layer, we use a Long-Short Term Memory (LSTM) \cite{lstm} to obtain a vector representation from the entire text. Finally, the output of this layer is connected to a dense layer to perform the text classification. This architecture uses a Dropout layer where a fraction of the input data is randomly assigned to zero in each training update phase to avoid overfitting. The block diagram for this architecture is shown in Figure \ref{fig:arquitecturaClasificador}.

\begin{figure}[ht!]
    \centering
    \includegraphics[width=4.0cm]{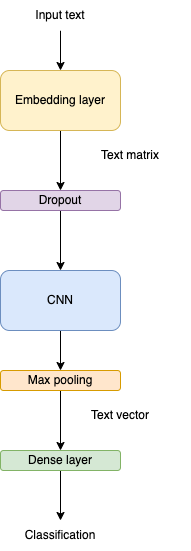}
    \caption{Block diagram for the proposed convolutional neural network architecture}
    \label{fig:arquitecturaClasificador_cnn}
\end{figure}

The second architecture is the same as the first one except for the Long-Short Term Memory layer; instead of this recurrent layer, we use a convolutional network followed by a pooling layer to get a vector representation for the entire text. Figure \ref{fig:arquitecturaClasificador_cnn} shows the block diagram for this architecture.


We compute the weight of each class in order to use a weighted cross-entropy as a loss function for both architectures. We use 10 epochs and batch size of 32 and \textit{K}-fold cross-validation with $K=5$ for all the experiments. For both cases, we used the Keras library \cite{chollet2015keras} to implement the network.

\subsection{Comparison experiments}\label{subsec:Comparacion}
In order to compare the performance of the proposed architectures, we carried out additional experiments in which the traditional machine learning algorithms were used: Support Vector Machines, Naive-Bayes, Logistic Regression, and Random forest. 
We use two different inputs to those algorithms. First, we obtain a text matrix using the pre-trained word2vec model trained on the Spanish Billion Word Corpus \cite{cardellinoSBWCE}, then we compute an average vector of each word on a text to obtain a vector representation for the whole text that we use as inputs to the listed algorithms above. We use the second input of a text vector representation obtained using a bag of words model.  

Similarly, the Facebook library FastText\cite{fastText} was used, which offers a tool that uses the ideas of \textit{Mikolov et al.} \cite{wordRepresentations} of efficient learning of word representations to train a linear classifier using a range constraint and a fast loss approximation. We also use the framework developed by \textit{Socher et al.,} \cite{autoencoders} which is based on recursive auto-encoders for sentence-level prediction.

\section{Results}\label{sec:Resultados}

\begin{table*}[ht!]
    \caption{Results with proposed architectures}
    \centering
    \begin{tabular}{|c|c|c|c|c|}
      \hline
    LSTM & Accuracy & Precision & Recall & F1 \\
      \hline
      Original dataset & 0.59 & 0.46 & 0.54 & 0.48\\
      SMOTE & 0.47 & 0.37 & 0.47 & 0.38\\
      Geometric-SMOTE & 0.42 & 0.33 & 0.41 & 0.32\\
      ADASYN & 0.39 & 0.34 & 0.40 & 0.33\\
      \hline
       CNN & Accuracy & Precision & Recall & F1 \\
       \hline
       Original dataset & 0.63 & 0.54 & 0.55 & 0.53\\
       SMOTE & 0.61 & 0.50 & 0.56 & 0.51\\
       \textbf{Geometric-SMOTE} & \textbf{0.64} & \textbf{0.54} & \textbf{0.56} & \textbf{0.54}\\
       ADASYN & 0.62 & 0.52 & 0.55 & 0.52\\
      \hline
    \end{tabular}
    \label{tab:results_neuralNets}
\end{table*}


Table \ref{tab:results_neuralNets} shows the results obtained with the deep learning architectures described in section \ref{subsec:Clasificador}. We can observe that the convolutional neural network architecture outperforms the recursive neural network architecture both when oversampling techniques are used and when the original dataset (with no oversampling techniques applied) is used.

In the case of the recurrent network, it obtains the best performance when no oversampling algorithms are used with 0.59 for Accuracy and 0.46, 0.54, and 0.48 for precision, recall, and F1, respectively. The best-achieved result using an oversampling technique is with the SMOTE algorithm. However, there is about a 10\% difference in terms of performance metrics with respect to results using the original dataset.

For the convolutional network, using the Geometric-SMOTE algorithm obtains the best performance with 0.64 of accuracy, 0.54 of precision, 0.56 of recall, and 0.54 of F1, which represent an improvement of 1\% in Accuracy, recall, and F1 when no oversampling techniques are used.

Table \ref{tab:resultadosComparacion_w2v} shows the results of the experiments using the classical machine learning algorithms with the average of the word vectors using the word2vec model as inputs to the classifiers as well as the results using FastText and RecursiveNN \cite{autoencoders}. Support Vector machines outperform all other algorithms in any experimental setup and also outperform the recurrent network in any case. The Naive-Bayes classifier obtains the worst performance of all the classifiers. In addition, the oversampling techniques do not represent an improvement for this classifier since the performance remains constant.

\begin{table*}[t]
    \caption{Results of comparison experiments using average word2vec vectors as input}
    \centering
    \begin{tabular}{|c|c|c|c|c|}
      \hline
    SVM & Accuracy & Precision & Recall & F1 \\
      \hline
      Original dataset & 0.65 & 0.61 & 0.45 & 0.49\\
      SMOTE & 0.64 & 0.59 & 0.47 & 0.49\\
      Geometric-SMOTE & 0.64 & 0.58 & 0.47 & 0.48\\
      ADASYN & 0.64 & 0.59 & 0.47 & 0.49\\
      \hline
       Naive-Bayes & Accuracy & Precision & Recall & F1 \\
       \hline
       Original dataset & 0.35 & 0.35 & 0.29 & 0.22\\
       SMOTE & 0.35 & 0.35 & 0.29 & 0.22\\
       Geometric-SMOTE & 0.35 & 0.35 & 0.29 & 0.22\\
       ADASYN & 0.35 & 0.35 & 0.29 & 0.22\\
      \hline
       Logistic Regression & Accuracy & Precision & Recall & F1\\
       \hline
       Original dataset & 0.62 & 0.56 & 0.37 & 0.40\\
       SMOTE & 0.61 & 0.52 & 0.39 & 0.40\\
       Geometric-SMOTE & 0.60 & 0.53 & 0.38 & 0.39\\
       ADASYN & 0.61 & 0.54 & 0.39 & 0.40\\
       \hline
       Random forest & Accuracy & Precision & Recall & F1 \\
       \hline
       Original dataset & 0.59 & 0.67 & 0.34 & 0.40\\
       SMOTE & 0.60 & 0.68 & 0.35 & 0.41\\
       Geometric-SMOTE & 0.59 & 0.68 & 0.36 & 0.39\\
       ADASYN & 0.69 & 0.68 & 0.34 & 0.40\\
       \hline
       FastText & Accuracy & Precision & Recall & F1 \\
       \hline
        - & 0.60 & 0.36 & 0.27 & 0.31 \\
       \hline
       RecursiveNN (Socher et al) & Accuracy & Precision & Recall & F1 \\
       \hline
       - & 0.12 & 0.09 & 0.11 & 0.10 \\
       \hline
    \end{tabular}
    \label{tab:resultadosComparacion_w2v}
\end{table*}

\begin{table*}[t]
    \caption{Results of comparison experiments using bag-of-words vectors as input}
    \centering
    \begin{tabular}{|c|c|c|c|c|}
      \hline
    SVM & Accuracy & Precision & Recall & F1 \\
      \hline
      Original dataset & 0.63 & 0.53 & 0.50 & 0.51\\
      SMOTE & 0.63 & 0.52 & 0.50 & 0.51\\
      Geometric-SMOTE & 0.63 & 0.53 & 0.51 & 0.51\\
      ADASYN & 0.63 & 0.52 & 0.50 & 0.51\\
      \hline
       Naive-Bayes & Accuracy & Precision & Recall & F1 \\
       \hline
       Original dataset & 0.58 & 0.60 & 0.28 & 0.32\\
       SMOTE & 0.58 & 0.60 & 0.29 & 0.33\\
       Geometric-SMOTE & 0.58 & 0.60 & 0.29 & 0.32\\
       ADASYN & 0.58 & 0.60 & 0.29 & 0.33\\
      \hline
       Logistic Regression & Accuracy & Precision & Recall & F1\\
       \hline
       \textbf{Original dataset} & \textbf{0.66} & \textbf{0.59} & \textbf{0.52} & \textbf{0.55}\\
       SMOTE & 0.66 & 0.58 & 0.52 & 0.54\\
       Geometric-SMOTE & 0.66 & 0.59 & 0.52 & 0.54\\
       ADASYN & 0.66 & 0.57 & 0.52 & 0.54\\
       \hline
       Random forest & Accuracy & Precision & Recall & F1 \\
       \hline
       Original dataset & 0.66 & 0.67 & 0.44 & 0.50\\
       SMOTE & 0.62 & 0.66 & 0.39 & 0.45\\
       Geometric-SMOTE & 0.66 & 0.69 & 0.44 & 0.51\\
       ADASYN & 0.63 & 0.66 & 0.39 & 0.45\\
       \hline
    \end{tabular}
    \label{tab:resultadosComparacion_bow}
\end{table*}

Finally, Table \ref{tab:resultadosComparacion_bow} shows the results of the experiments with the classical machine learning algorithms using bag-of-words vectors as input. The best performing algorithm is Logistic Regression, followed by Support Vectors Machine, Random Forest, and Naive-Bayes. In this case, the oversampling techniques do not represent an improvement in the classifier's performance since the best result is obtained using the original dataset. Furthermore, this is the best result of all experimental setups; it outperforms the convolutional neural network when Geometric-SMOTE is used as an oversampling technique.

\section{Conclusions}\label{sec:conclusiones}
In this article, we present an automatic classifier of job offers using two neural network architectures to improve the operation of the DiCoDe system in Mexico City. For this classification task, we face the class imbalance problem in the field of multiclass classification. To tackle this problem, 3 synthetic oversampling algorithms were used, SMOTE, Geometric-SMOTE, and ADASYN. The results of the proposed architectures were compared with the classic machine learning algorithms Support Vector Machine, Naive-Bayes, Logistic Regression, and Random forest using average word2vec vectors and bag-of-words vectors as input as well as the FastText library and a framework based on recursive auto-encoders. 

The convolutional neural network architecture outperforms the recurrent neural network in all experimental setups. It also outperforms the FastText library, the recursive auto-encoder-based framework, and all machine learning algorithms when average word2vec vectors are used as inputs. The best result for this architecture is achieved when the Geometric-SMOTE algorithm is used as an oversampling method.

Nevertheless, our proposed convolutional neural network is outperformed by a logistic regression classifier when no oversampling techniques were used by 0.02 in Accuracy, 0.04 in precision, and 0.01 in F1.

Given the characteristics of our neural network architecture, future steps examine the use of a hybrid architecture that includes recurrent and convolutional neural networks through a \cite{CRAN} attention mechanism. On the other hand, in relation to oversampling methods, the effects of using methods based on Generative Adversarial Networks can be studied. These networks seek to approximate the real distribution of the data of the minority classes instead of using local information with the k-nearest neighbor algorithm \cite{GANs}.



\section*{Acknowledgment}
The work was done with partial support from the Mexican Government through the grants A1-S-47854 and CB A1-S-27780 of the CONACYT-Mexico, grants of  PAPIIT-UNAM projects TA400121 and TA101722, and grants 20211784, 20211884, and 20211178 of the Secretar\'ia de Investigaci\'on y Posgrado of the Instituto Polit\'ecnico Nacional, Mexico. The authors also thank the CONACYT for the computing resources brought to them through the Plataforma de Aprendizaje Profundo para Tecnolog\'ias del Lenguaje of the Laboratorio de Superc\'omputo of the INAOE, Mexico. The authors also acknowledge the support of the DiCoDe project.

%
%
%
\bibliographystyle{splncs04}
\bibliography{reference}
%




\end{document}